\documentclass{ws-ijmpa}

\renewcommand\gg{$\gamma \gamma$\space}
\newcommand\hgg{$h \to \gamma \gamma$\space}
\newcommand\dz{D\O \space}

\begin{document}

\markboth{Alex Melnitchouk}
{Search for non-SM Light Higgs Boson in the \hgg Channel
at \dz in Run II}

\catchline{}{}{}{}{}
%
%%%%%%%%%%%%%%%%%%%%%%%%%%%%%%%%%%%%%%%%%%%%%%%%%%%%%%%%%%%%%%%%%%%%

\title{Search for non-SM Light Higgs Boson in the \hgg Channel
at \dz in Run II}

\author{\footnotesize ALEX MELNITCHOUK\footnote{
Fermilab, MS\#352, Batavia, IL, 60510, USA.}}

\address{Department of Physics and Astronomy,\\ The University of Mississippi,\\ 
108 Lewis Hall, P.O. Box 1848\\
University, MS 38677-1848
}

\maketitle

\pub{Received (31 October 2004)}{Revised (31 October 2004)}

\begin{abstract}
A search for non-SM light Higgs boson with an enhanced branching fraction
into photons in $p \bar{p}$ collisions
at the Fermilab Tevatron is presented 
using Run II D\O\ data taken between April 2002 and September 2003.
We set 95\% CL limits on the diphoton branching fraction as a function
of Higgs mass for Fermiophobic and Topcolor Higgs scenarios. 

\end{abstract}

\section{Motivation}	%) A SECTION HEADING

$\gamma \gamma$ is a very clean signature which
makes it promising and important especially for the hadron collider environment.
There are many extensions of the SM that allow enhanced $\gamma \gamma$ decay rate
of the Higgs largely due to suppressed couplings with fermions
\cite{SteveM}. We consider Fermiophobic Higgs (no Higgs-fermion couplings)
and Topcolor Higgs (coupling with the top quark is allowed) 

\section{Dataset}

We used the data collected by the \dz detector between 
April 2002 and September 2003.
The integrated luminosity of this sample is $191.0 \pm 12.4 \: {\rm pb}^{-1}$
Trigger selection was done with high $p_{T}$ di-EM\footnote{EM stands 
for ``electromagnetic object", i.e. electron or photon} trigger.
Offline we require two reconstructed photon objects in the good
$\eta$ fiducial region ($|\eta|<1.05$, $1.5<|\eta|<2.4$)
with $p_{T} > 25 \: \mbox{GeV}$. We require that the $p_{T}$ of the diphoton system
is above 35 GeV \cite{mythesis}.

\section{Backgrounds}
Major sources of background to $h \to \gamma \gamma$ are 
$Z/\gamma^* \to ee$, direct diphotons, $\gamma  +jet$, and multijet QCD processes.

\section{\label{sec:level1_Mass_and_Event_Counts} Results}
\begin{figure}[tbhp]
\centering
\includegraphics[scale=0.5]{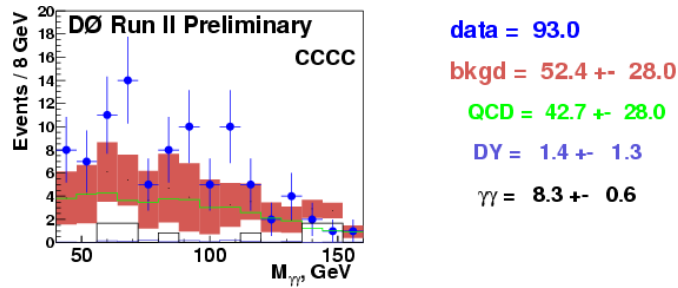}
%%\epsfxsize=6.5in
%%\epsffile{../FINALFINAL.022904closerllook.diphpt.gt.35.eps}
\caption{
\gg invariant mass distributions and event counts for different event topologies
with new photon ID after analysis optimization cut ($p_{T}^{\gamma\gamma} > 35 \: \mbox{GeV}$) 
is applied. Points -- \gg spectrum observed in data, 
red rectangles -- total SM background with errors,
green line -- QCD background, brown line -- Drell-Yan background,
black line -- direct diphoton background.}
\label{fig:massspectrumwithdiphptgt35_newphotonID}
\end{figure}

\begin{figure}[tbhp]
\centering
\includegraphics[scale=0.5]{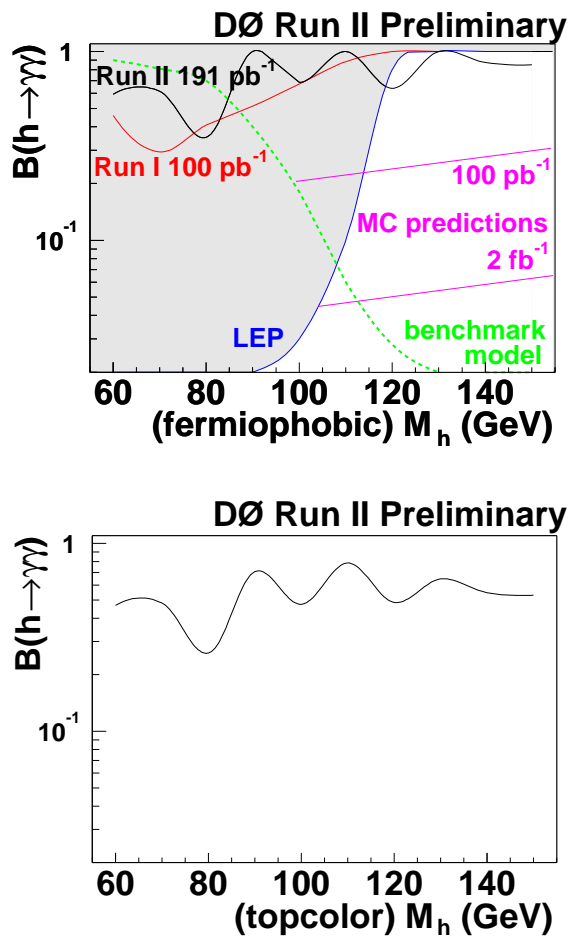} 
\caption{
95 \% 
CL limits on the Higgs decay branching fraction into photons
as a function of mass (black curve).
Top plot -- fermiophobic Higgs scenario, bottom plot -- topcolor Higgs scenario.
On the top plot exclusion contours from 
D\O \space Run I 
\protect\cite{D0_hgg} 
(red) and  LEP \protect\cite{LEP_Higgs} (blue) are overlaid.
Magenta lines show $100 \: {\rm pb}^{-1}$ and $2 \: {\rm fb}^{-1}$ 
Monte Carlo predictions for Tevatron Run II \protect\cite{Greg_paper} based
on Run I \dz and CDF efficiencies and misID rates.
Green points -- theoretical curve  
for benchmark fermiophobic Higgs model \protect\cite{LEP_Higgs}.
}
\label{fig:limits_newphotonID}
\end{figure}

The $\gamma \gamma$ invariant mass distributions for the data and predicted background,
as well as the event yields are shown in Fig. \ref{fig:massspectrumwithdiphptgt35_newphotonID}.
Figure \ref{fig:limits_newphotonID} shows 95\% CL limits on the B(\hgg)
as a function of Higgs mass. Comparison is made with \dz Run I and LEP results
as well as Run II Monte Carlo studies for Tevatron \cite{Greg_paper}.  
\section{\label{sec:level1_conclusions} Conclusions}
A search for non-SM light Higgs boson with an enhanced branching fraction
into photons was performed using $191 \: {\rm pb}^{-1}$ of data collected
by the D\O\ experiment in Run II of the Fermilab Tevatron.
We set 95\% CL limits on the diphoton branching fraction as a function
of Higgs mass for Fermiophobic and Topcolor Higgs scenarios.
Our current sensitivity is comparable with Run I.
\section*{Acknowledgments}
We thank the staffs at Fermilab and collaborating institutions, 
and acknowledge support from the 
Department of Energy and National Science Foundation (USA),  
Commissariat  \` a L'Energie Atomique and 
CNRS/Institut National de Physique Nucl\'eaire et 
de Physique des Particules (France), 
Ministry for Science and Technology and Ministry for Atomic 
   Energy (Russia),
CAPES, CNPq and FAPERJ (Brazil),
Departments of Atomic Energy and Science and Education (India),
Colciencias (Colombia),
CONACyT (Mexico),
Ministry of Education and KOSEF (Korea),
CONICET and UBACyT (Argentina),
The Foundation for Fundamental Research on Matter (The Netherlands),
PPARC (United Kingdom),
Ministry of Education (Czech Republic),
A.P.~Sloan Foundation,
Civilian Research and Development Foundation,
Research Corporation,
Texas Advanced Research Program,
and the Alexander von Humboldt Foundation.
\clearpage


\begin{thebibliography}{0}
\bibitem{SteveM}
         S. Mrenna, J. Wells, Phys. Rev. {\bf D63}, 015006 (2001).

\bibitem{mythesis} Alex Melnitchouk,Ph.D Thesis, 
                   Brown University (2003) (Unpublished).


\bibitem{D0_hgg} B. Abbot {\it et al.} (\dz Collaboration),
                 Phys Rev. Lett. {\bf 82}, 2244 (1999);
                 Brian Lauer,Ph.D Thesis, 
                 Iowa State University (Unpublished)(1997).

\bibitem{LEP_Higgs} A. Rosca, hep-ex/0212038 (2002).

\bibitem{Greg_paper} G. Landsberg, K. Matchev, Phys. Rev. {\bf D62}, 
                     035004 (2000).

\end{thebibliography}
\end{document}